# Complexity-Scalable Direct Geolocation and Cancellation of Terrestrial GNSS Jammers: Single-Satellite and Multi-Antenna Experiments in Low Earth Orbit

Giacomo Pojani, Javier Tegedor, Joaquim Fortuny-Guasch and Francesco Menzione,
*European Commission, Joint Research Centre*
David Evans and Tim Oerther, *European Space Agency*
Maximilian Henkel, *Technical University of Graz*
Jonas Stene Lindbjør, *Testnor*

## BIOGRAPHIES

Giacomo Pojani earned his M.Sc. and Ph.D. in Telecommunications Engineering from the University of Bologna, Italy. In 2017, he was visiting researcher at the European Space Research and Technology Centre (ESTEC) with focus on GNSS interference mitigation and geolocation. He then worked across the semiconductor, defense, and space-based intelligence industries for several years, before joining the Joint Research Centre (JRC) of the European Commission (EC) in 2025, where he currently serves as scientific officer for GNSS resilience and security.

Javier Tegedor holds a M.Sc. in Telecommunications Engineering at Polytechnic University of Valencia, Spain, and a Ph.D. in Geomatics from the Norwegian University of Life Sciences. He has over 20 years of experience in GNSS in both private and public sectors, including the European Space Operations Centre (ESOC). He is currently working as scientific officer at the EC's JRC, providing technical support to programs of positioning navigation and timing (PNT) as well as GNSS resilience and security.

Joaquim Fortuny-Guasch obtained his M.Sc. in Telecommunications Engineering from the Technical University of Catalonia, Spain, and Pd.D. from the University of Karlsruhe, Germany. Since 1993 he has worked for the JRC of the EC, where he serves as a senior scientific officer. He heads the unit sector specializing in GNSS resilience and security and is currently working on antenna characterization, GNSS interference monitoring and laboratory testing of GNSS receivers.

Francesco Menzione received his M.Sc. and Ph.D. in Aerospace Engineering and Satellite Navigation from the University of Naples Federico II, Italy. From 2012 to 2021, he worked in the aerospace industry as navigation and control engineer. In 2021, he joined the EC's JRC, where he has been contributing as scientific officer to research initiatives in the realm of low Earth orbit (LEO) PNT, space service volume (SSV), non-terrestrial networks (NTN), GNSS interference monitoring, and GNSS-based remote sensing.

David Evans started his career in 1992 at ESOC. He then joined EUTELSAT, eventually becoming the manager of the Satellite Control Centre. He returned to ESOC in 2007 and initiated the OPS-SAT mission, the first CubeSat to controlled by the European Space Agency (ESA), which spawned follow-on missions. He has four ESA patents on telemetry compression, one of which was adopted as a CCSDS international standard.

Tim Oerther received his M.Sc. in Electronics Engineering and Information Technology from the Karlsruhe Institute of Technology. He is currently a spacecraft operations engineer at ESA. In 2023, he joined the OPS-SAT Space Lab, contributing to OPS-SAT-1 flight operations and the preparation of future missions. He was responsible for experimental services across OPS-SAT missions, including the OPS-SAT PRETTY satellite.

Maximilian Henkel received his M.Sc. in Information and Communication Technology from Technical University (TU) of Graz, Austria. He then worked as a research engineer for ESA's OPS-SAT-1 mission, focusing on its processing platform. After the launch in 2019, he supported the entire mission operations and successfully used the experience on reconfiguring the on-board FPGA to earn his Ph.D. in 2024. He is currently a member of ESA's OPS-SAT Space Lab and the OPS-SAT PRETTY operations team.

Jonas Stene Lindbjør received his M.Sc. from the Norwegian University of Science and Technology, Norway. He is the chief technology officer (CTO) of Testnor, a Norwegian company specializing in large scale open-air and laboratory testing of radio-frequency interference.

## ABSTRACT

Monitoring the radio-frequency (RF) spectrum from space imposes demanding requirements to satellite platforms in terms of communication bandwidth and computational resources, which are necessary for the downlink, the storage, and the processing of high-throughput I/Q samples. This paper analyzes in depth the quasi-direct geolocation (QDG) as a technique to enable the exploitation of satellites of opportunity in low Earth orbit (LEO) to sense the spectrum in the bands of global navigation satellite systems (GNSS). This is a technique of passive RF geolocation and consists of an ensemble of signal processing algorithms, which compress the I/Q samples and process the compressed data through fast delay-Doppler shift matching and interferometry in a quantized time-frequency domain. These algorithms speed up the exhaustive search of multiple RF sources in the position domain. The efficiency gain addresses the bottleneck that prevents the employment of satellites, which are limited in downlink capacity and on-board computational power. These satellites are usually constrained in size, weight and power (SWaP) and represent most of the spacecrafts in LEO. The ability to exploit assets as such for the geolocation of terrestrial GNSS jammers in near real time is instrumental the performance of a multi-constellation GNSS RFI monitoring system.
The present work describes the mathematical framework and precision bounds, introduces single- and multi-antenna uses cases, combines different compression methods, and evaluates the geolocation accuracy with real data. The I/Q samples were collected by a repurposed GNSS reflectometry (GNSS-R) satellite, OPS-SAT PRETTY, in a dedicated test session during Jammertest 2025. The experimental results demonstrate the capability to geolocate GNSS jammers with different signal-to-noise ratios (SNR) with extremely high compression ratios.

## 1. INTRODUCTION

Every global navigation satellite system (GNSS) has a space segment consisting in a constellation of satellites, which broadcast synchronized signals that are modulated by orbit and clock data. By acquiring, tracking, and demodulating these signals, user GNSS receivers worldwide can solve for their own position, velocity, and time (PVT) with high precision. Because these signals are wideband and, at the time being, mostly transmitted from satellites in medium Earth orbit (MEO), they arrive at the Earth's surface with extremely weak power spectral densities. These densities are below the noise floor on ground. This makes the reception of GNSS signals vulnerable to stronger RF interference (RFI) coming from terrestrial sources, including intentional ones, such as jammers, meaconers and spoofers. Among them, jammers are typically the most powerful interferers. In fact, jamming is a denial-of-service attack carried out by overpowering the GNSS signals in the surroundings, where the victim GNSS receivers suffer from saturated RF front ends or fail to acquire, track, and demodulate signals from enough satellites. These types of attacks can be performed simultaneously across multiple frequency bands to disrupt several systems beyond GNSSs. The increasing occurrence of jamming inside and near the GNSS bands has motivated the development of technologies to mitigate the interference picked up by user receivers as well as to detect and geolocate the sources of interference.

Satellites in low Earth orbit (LEO) can be both affected by GNSS RFI and exploited for monitoring GNSS RFI over various bands. In fact, most LEO satellites are equipped with GNSS receivers for various purposes: positioning navigation and timing (PNT), precise orbit determination (POD), GNSS radio occultation (GNSS-RO), GNSS reflectometry (GNSS-R), etc. Therefore, these receivers can act as GNSS performance sensors in space: while the satellites fly over GNSS jammers, various metrics related to the GNSS signals received and the PVT solutions calculated by the receiver exhibit degradations that can be used to detect regions affected by RFI. Beyond detection, to further enable the characterization and geolocation of RFI sources within such regions, the satellites shall act as GNSS spectrum sensors. This requires satellites capable of recording synchronized snapshots of I/Q samples in space. The capability of sensing the GNSS spectrum from LEO offers a few unique advantages. First, the stand-off distance allows for the reception of RFI with low-probability saturation and good sensitivity via nadir-facing antennas towards ground. The same distance also protects the reception of GNSS signals through zenith-facing antennas towards MEO. Secondly, the satellite field view can cover large swaths of the Earth's surface with short and regular revisit times. Lastly, the propagation effects of the received signals are traceable to the line-of-sight paths with predictable atmospheric effects.

Characterizing and geolocating RFI sources visible from space implies the recording and processing of synchronized snapshots of I/Q samples, which carry much more information than metrics or proxy indices of GNSS performance. However, this processing is carried out through computationally intensive algorithms. Therefore, normally, the recorded I/Q samples shall be stored and relayed to ground, where they are then post-processed with powerful computers. This requirement challenges the inherent limitations of platforms in space: I/Q samples are high-volume data with rates spanning from megabytes to gigabytes per second, which stress the limits of the on-board storage and the downlink capacity. To cope with these limitations, there are two options: either compressing the data before downlink or processing the data in orbit through on-board computers (OBC).

From a broader perspective, any GNSS RFI monitoring system would benefit from integrating as many LEO satellites as possible. In fact, this is instrumental to fulfil requirements in terms of coverage, availability, and accuracy of the RFI geolocation. However, in the current LEO landscape, most space assets are single satellites that fly on platforms that are characterized by low size, weight, and power (SWaP). Cooperative fleets, including tandems, clusters, or the synchronized conjunction of multiple satellite passes, remain relatively rare. Therefore, the effective exploitation of single SWaP-constrained satellites is crucial to the performance of space-based RFI monitoring systems over wide geographical regions, even if such platforms do not feature high-speed downlinks or powerful OBCs. Addressing the requirements imposed by RFI geolocation from LEO can increase the number of current and future satellites that can participate to a multi-constellation RFI monitoring system. For example, GNSS spectrum sensors can be integrated as hosted payloads within emerging non-terrestrial network (NTN) infrastructure layers.

The authors are introducing in (Pojani et al., 2026) a technique for the low-complexity direct geolocation of RF emitters, which is named quasi-direct geolocation (QDG). The work in this paper explores several performance factors of QDG when applied to geolocate GNSS jammers from LEO in a single-satellite scenario with multiple antennas. This technique comprises an ensemble of algorithms. On the one hand, they can be used to compress the I/Q samples and to relay to ground only the data with minimal signal information. On the other hand, they can process these compressed data in orbit with low-power OBCs, thus bypassing the need for downlink. The performance of QDG is assessed experimentally with the I/Q samples collected in (Tegedor et al., 2026) during Jammertest 2025 by a 3U CubeSat in LEO: OPS-SAT PRETTY. The CubeSat was designed for GNSS-R and later repurposed after the completion of its original scientific mission, as documented in (Zeif et al., 2020; Pirat et. al. 2024).
Another original contribution of this paper is the investigation of methods to cancel the signals received. After the cancellation of the powerful test jammer, other distant and weaker in-band emissions stand out. These emissions are traced to regions that are famously affected by persistent and high-power RFI in L-band, as evidenced by the irregularities in the NIC (Navigation Integrity Category) registered by the avionic GNSS receivers on local aircrafts (Lo et al., 2026). The NIC is a GNSS performance proxy index contained in messages from the automatic dependent surveillance broadcasts (ADS-B) and used to monitor air traffic.

## 2. STATE OF THE ART OF RADIO-FREQUENCY GEOLOCATION

Passive geolocation of RF emitters is the process of determining the geographic coordinates of signal sources using one or more passive receivers. Traditional techniques generally comprise two estimation steps and are used in a number of works (Tegedor et al., 2026; CaJacob et al., 2016; Pojani et al., 2018; Ellis et al., 2020, 2022; Murrian et al., 2021; Menzione et al., 2025; Clements et al., 2026). In the first step, they measure physical observables of the received signals, such as times of arrival (TOA), frequency of arrival (FOA), angles of arrival (AOA), etc., as well as inter-receiver differences of these observables, such as time differences of arrival (TDOA), frequency differences of arrival (FDOA), etc. These time-frequency measurements are estimated from the I/Q samples collected by the receivers and then they are aggregated into time series. In the second step, the time series of measurements are fed to nonlinear estimators, such as Gauss-Newton optimization and Kalman filtering, to geolocate a given number of emitters. These estimators iteratively or recursively minimize a cost function, which is defined based on the error between observed and predicted measurements.
The two-step techniques suffer from two main issues. First, they are inherently sub-optimal because they split the estimation problem into two sub-problems or steps. In the first step, the signal information carried by the I/Q samples collapse into intermediate and independent measurements. Consequently, the accuracy and sensitivity of the estimation tend to degrade. Secondly, their performance is dependent on the correct aggregation of a multitude of measurements of TOA, FOA, TDOA, FDOA, AOA, etc. into distinct sets of time series across the receivers, where each set of series shall correspond to one individual emitter. This aggregation process is challenging in the presence of multiple emitters with a variety of modulations and power levels, especially if the number of emissions is unknown and changes over time. In fact, ambiguities and artefacts arise in this domain from the overlapping of the received signals as well as the correlations both inter-signal and intra-signal. They occur especially with observation of large numbers of uncooperative emitters, which, remarkably, is the case of RFI sources visible from LEO in the GNSS bands. These challenges pose the reliability of the estimates at stake.

To overcome the shortcomings of traditional techniques, single-step techniques were proposed in (Amar et. al, 2008; Weiss, 2011) to geolocate emitters from multiple receivers. The recent publication in (Clements et al., 2023) demonstrated the application of these techniques to geolocate GNSS RFI sources from two LEO satellites. These techniques are collectively referred to as direct geolocation (DG) or direct position determination (DPD), because they perform an exhaustive search directly in the position domain. The search is based on cross-correlating the I/Q samples over predicted observables, such as TDOA/FDOA time series, which are parametrized for a grid of candidate positions. In this domain, individual emitters are separated by their physical distance, and the maximum-likelihood estimates of their geographic coordinates can be found at the candidate positions, where the cross-correlation magnitude exhibits significant peaks. This process bypasses the need for intermediate measurements and measurement aggregation, and it requires no prior knowledge of the number of visible emitters. By working with the full signal information in the I/Q samples, DG/DPD can outperform two-step techniques in terms of accuracy sensitivity, especially with low signal-to-noise ratio (SNR) and/or short integration times. However, this improvement comes with two disadvantages. The first issue is a simplifying assumption: the time series of observables are parametrized for an emitter that is stationary within the integration. The second is an issue of complexity due to the brute-force nature of exhaustive searches: a search space covering a wide geographic area means cross-correlating I/Q samples over grids made by hundreds of thousands of candidate positions. In addition, generally, the longer the integration time is or the wider the emission bandwidth is, the finer shall be the grid. Consequently, DG/DPD imposes a heavy computational burden, which has confined its implementations to the post-processing of I/Q samples with graphics processing units (GPU) as in (Clements et al., 2025).

In the context of GNSS RFI monitoring from space, powerful GPUs are generally not available on SWaP-constrained satellites. They are rather integrated into high-performance computers. Therefore, DG/DPD is hardly feasible or practical in orbit, especially with platforms limited in SWaP. Consequently, these techniques entail that massive volumes of I/Q samples, on the order of tens of GB, need to be relayed to ground from multiple satellites in LEO. This downlink introduces a significant latency, especially with few ground stations operating in S- and C-band. This latency can be reduced only by restricting the collection to areas or time slots of interest, by splitting each collection into short-time acquisitions, by deploying an extensive network of ground stations, and by upgrading to high-capacity downlinks in X-, Ku-, or Ka-band or in near-infrared (NIR).

The technique of QDG aims at minimizing the computational burden and time to geolocate multiple RF emitters with fast exhaustive searches even on large grids of candidate positions. This technique pre-preprocesses the I/Q samples through a discrete and linear transform, which is used to quantize and adaptively compress the signal information in the time-frequency domain. The compression ratio is controllable. As a result, the QDG is expected to improve over DG/DPD in terms of speed and flexibility, by trading accuracy and sensitivity for reduced complexity. However, it also restricts the classes of signal sources that can be geolocated, where the restriction depends on the specific time-frequency transform and compression algorithms used. For example, as argued in (Pojani et al., 2026), QDG is supposed to effectively geolocate emitters of modulations that are instantaneously narrowband: either stationary, such as Continuous Waves (CW), or frequency-swept, such as periodic chirps and frequency-hopped CWs. These are most common signals broadcasted by GNSS jammers inside GNSS bands. Therefore, under specific restrictions, QDG can be used as low-complexity and degraded-accuracy alternative to DG/DPD.

## 3. JAMMERTEST AND OPS-SAT PRETTY

Jammertest is regarded as the largest open-air GNSS RFI exercise in the world (Gerrard et al., 2026). It is organized annually in Andøya, Norway, with the coordination of Testnor and several Norwegian public authorities. The added value of this exercise is the detailed knowledge of the RFI environment, i.e., position, power, frequency, waveform, etc. of all test emitters. Beyond these exercises, most of real-world scenarios outdoors are illegal and/or military RFI events, in which the ground truth is either unknown or only partially disclosed.
The data examined in this paper comprise I/Q samples collected during Jammertest 2025 as part of an experiment coordinated by the European Space Agency (ESA) OPS-SAT Space Lab initiative. The Lab provides cost-free, quick access to different spacecrafts for in-flight experimentation for any European or Canadian entity. The Austrian 3U CubeSat OPS-SAT PRETTY, which stands for Passive REflecTometry and dosimeTrY, joined the initiative in 2025 after successfully completing its original scientific mission of GNSS-R in-orbit demonstration in (Pirat et al., 2024).
The heart of the spacecraft is the satellite experimental processing platform (SEPP), which is used as a payload OBC during nominal operation and offers high processing capacities for the payloads as well as space-to-ground communications in S-band. The software-defined radio (SDR) for the passive reflectometer payload is based on the AD9361 transceiver IC from Analog Devices. The RF front end is connected to two Earth-facing patch antennas, which are specifically tuned for GPS/Galileo L5/E5a frequency as per GNSS-R mission requirements. The satellite also includes a commercial off-the-shelf (COTS) GNSS receiver supporting GPS/Galileo L1/E1 tracking.

## 4. SINGLE-SATELLITE QUASI-DIRECT GEOLOCATION

This section details the mathematical framework for QDG in a scenario with a single satellite equipped with one- or two-element antennas. The reader may refer to (Pojani et al., 2026) for a multi-receiver generalization.

### 4.1 Signal Model

Let us adopt the standard analytic signal model with the narrowband approximation valid in GNSS bands. For simplicity and in line with the assumption underlying DG/DPD and QDG, let us assume that the signal is transmitted by a stationary emitter. This model over time $t$ is the complex envelope $s(t)$ upconverted to the transmit frequency $f_{TX}$ and baseband bandwidth $B_{TX}$:

$$x(t) = s(t)\, e^{j\,2\,\pi\, f_{TX}\, t}, \qquad f_{TX} \ll B_{TX} \tag{1}$$

Let us consider the reception and downconversion of this signal by a moving receiver. If we neglect noise, propagation effects, and the impact of the receiver RF front end, the ideal complex envelope in reception is:

$$y(t) = s\left(\left(t - \tau(t)\right)\left(1 - \beta(t)\right)\right) e^{j2\pi\left(\Delta f_{LO} t + \int_0^t f_d(\tau) d\tau\right)} \tag{2}$$

The TOA of this signal at the time $t$ is:

$$\tau(t) = \frac{\rho(t)}{c} + \delta t_{RX}(t) - \delta t_{TX}(t) \quad (3)$$

where the time of flight, which is equal to the geometric range $\rho(t)$ divided by the speed of light c, is offset by the receiver clock bias $\delta t_{RX}(t)$ and the transmit time $\delta t_{TX}(t)$. Given the matrix transpose operator $[\cdot]^T$, the range between the receiver position vector $\mathbf{p}_{RX}(t)$ and the static position vector $\mathbf{p}_{TX}$ is:

$$\rho(t) = \sqrt{\mathbf{r}^T(t)\mathbf{r}(t)} \quad (4)$$

with the range vector $\mathbf{r}(t) = \mathbf{p}_{RX}(t) - \mathbf{p}_{TX}$. The complex envelope $\mathbf{y}(t)$ is affected by the time dilation/compression factor related to the TOA as $\beta(t) = d\tau(t)/dt$. If the frequency $f_{RX}$ driven by the receiver local oscillator (LO) does not coincide with $f_{TX}$, the received signal FOA is off by a constant frequency bias $\Delta f_{LO} = f_{TX} - f_{RX}$. The time-varying component of this FOA at time t is:

$$f_d(t) = -f_{TX}\,\beta(t) = -\frac{\hat{\mathbf{r}}^T(t)\,\mathbf{v}(t)}{\lambda} - f_{TX}\left(\delta\dot{t}_{RX}(t) - \delta\dot{t}_{TX}(t)\left(1 - \delta\dot{t}_{RX}(t)\right)\right) \quad (5)$$

where $\lambda = c/f_{TX}$ is the carrier wavelength, and the physical Doppler shift is proportional to the receiver velocity vector $\mathbf{v}(t)$ that is projected along the range unit vector $\hat{\mathbf{r}}(t) = \mathbf{r}(t)/\rho(t)$, while $\delta\dot{t}_{RX}(t)$ and $\delta\dot{t}_{TX}(t)$ are the clock drifts of the receiver and the emitter, respectively.

Let us add an antenna array to the receiver. The received signal model in (2) can be applied to each element of the array. Without loss of generality and in conformity with the experimental setup, we consider an array of two elements, which are indexed by 0 and 1, and placed at ends of a baseline vector $\mathbf{d}(t)$. Their position vectors are $\mathbf{p}_{RX,i}(t) = \mathbf{p}_{RX}(t) \mp 0.5\,\mathbf{d}(t)$ with $i = 0, 1$, at vectors $\mathbf{r}_i(t) = \mathbf{p}_{RX,i}(t) - \mathbf{p}_{TX}$ and ranges $\rho_i(t)$, with $\mathbf{p}_{RX}(t)$ set in the middle of the array. Differencing simultaneous TOAs and FOAs at these two elements cancels out the transmit time $\delta t_{TX}(t)$, the emitter clock drift $\delta\dot{t}_{TX}(t)$, the receiver clock bias $\delta t_{RX}(t)$ and drift $\delta\dot{t}_{RX}(t)$, resulting in the following TDOAs and FDOAs:

$$\Delta\tau(t) = \frac{\rho_0(t)}{c} - \frac{\rho_1(t)}{c} \quad (6)$$

$$\Delta f_d(t) = -\frac{1}{\lambda}\left(\hat{\mathbf{r}}_0^T(t)\mathbf{v}_0(t) - \hat{\mathbf{r}}_1^T(t)\mathbf{v}_1(t)\right) \quad (7)$$

The phase difference of arrival (PDOA) between the two antenna elements of the signal may be mathematically expressed as a function of a AOA plus a constant phase offset $\Delta\phi_{RX} \in [-\pi,\ \pi[$, which is due to differences in the antennas and the downstream RF chains. The time-varying component of this PDOA at time t is strictly geometrical:

$$\Delta\phi(t) \approx \frac{2\pi\, d \cos\theta(t)}{\lambda} \quad (8)$$

where the approximation holds if the transmitter is in the very far field of the antennas, namely many wavelengths away from the receiver, with the AOA of the signal being expressed as:

$$\theta(t) = \cos^{-1}\left(\frac{-\mathbf{r}^T(t)\,\mathbf{d}(t)}{\rho(t)\,d}\right), \qquad d = |\mathbf{d}(t)| \leq \frac{\lambda}{2} \ll \rho(t) \quad (9)$$

### 4.2 Time-Frequency Transform

The QDG is an ensemble of algorithms, the first of which consists in pre-processing the I/Q samples through a discrete and linear time-frequency transform, such as the short-time Fourier transform (STFT) and the S-transform, which are reviewed in (Abdoush, 2019b). The STFT may be regarded as a special case of the S-transform, in which the trade-off between time and frequency resolutions is fixed by setting a constant analysis window. The STFT is simple and computational efficient: it can be processed either in batches with the fast Fourier transform (FFT) or sample by sample through a bank of digital filters.

Let us consider the ideal digitization of the noiseless received signal model in (2) is $y_i[k] = y_i(kT_s)$ at each element of the antenna array, with a sampling period $T_s = 1/f_s$. The discrete STFT of the samples $y_i[k]$ at the m-th time bin and n-th frequency bin for the i-th antenna is:

$$Y_i[m,n] = \sum_{k=0}^{N-1} y_i[k]\, w^*[k - m\,L] e^{-j2\pi \frac{n}{N} k} \tag{10}$$

where $w^*[k]$ is the complex conjugate of the analysis window $w[k]$ sliding over the samples, N is the size of this window coinciding with the number of frequency bins, and L is the hop size or the number of samples by which the window slides. The time resolution $\Delta t = L\, T_s$ and the frequency resolution $\Delta f = f_s/N$ are inversely proportional and fixed by the choice of $w[k]$. If we upper bound the hop size to half the window size, so that all samples in $y[k]$ enter (10), the number of time-frequency bins of the entire STFT is always greater than or equal to the number of signal samples:

$$N\,M = N\left(\left\lfloor \frac{K-N}{L} \right\rfloor + 1\right) \geq K, \qquad L \leq \left\lfloor \frac{N}{2} \right\rfloor \tag{11}$$

where K is the number of samples of $y[k]$, $k = 0, \ldots, K-1$, and the number M of time bins or hops or windows is derived without zero-padding to handle boundary windows. The integration time of the STFT is: $T = K\,T_s \sim (N + (M-1)L)\,T_s$.

**4.3 Noise Estimation**

The second algorithm for QDG is the noise estimator. There exist many algorithms for noise estimation, which are performed either recursively or in batches. In the present paper, we estimate noise by iteratively fitting a probability density function to the histogram of the real and imaginary parts of the STFT bins. This works by assuming that only a fraction of the STFT, which is used for noise estimation, contains bins with strong signal energy that skew the tail of the distribution.
Let us assume the standard white Gaussian noise model through the random process $n[k]$ that is fully characterized by the standard deviation $\sigma_n$: $z_i[k] = y_i[k] + n[k]$. Since the receiver is one and equipped with separate RF chains (see AD9361), the noise across the antennas is assumed as independent, identically distributed and characterized by the same statistical moments. Therefore, algorithm produces an estimate $\hat{\sigma}_n \approx \sigma_n$, that is common across the elements of the antenna array. From this estimate we can derive the SNR per time-frequency bin of the STFT of $y_i[k]$ as:

$$\mathrm{SNR}_i[m,n] = \frac{|Z_i[m,n]|^2 - 2\hat{\sigma}_n^2 E_w}{2\hat{\sigma}_n^2 E_w}, \qquad E_w = \sum_{k=0}^{N-1} |w[k]|^2 \tag{12}$$

where $Z_i[m,n]$ is the STFT of $z_i[k]$ and $E_w$ is the energy of the analysis window. Under the additive white Gaussian model, the squared magnitude of a STFT bin containing noise is modeled as a Chi-squared random variable. As proven in (Borio et al., 2008), it follows that we can define a binary decision test to determine the bins that contain meaningful power. Statistically, these bins have energy above the noise if they exceed the threshold $\alpha$ with arbitrary false-alarm rate $P_{FA}$, namely:

$$\alpha = -2\hat{\sigma}_n^2 E_w \log P_{FA} \tag{13}$$

**4.4 Two-Antenna Cross-Correlation Approximation**

The third and last algorithm of QDG consists in approximating the cross-correlation of I/Q samples among antennas in the position domain. Given two K-size arrays of samples (2) from the two elements of an antenna array, their cross-correlation at an arbitrary position vector $\mathbf{p}$ over the coherent integration time $T = K\,T_s$ is expressed as:

$$S(\mathbf{p}) = \sum_{k=0}^{K-1} y_0[k] y_1^*[k] e^{-j\Delta\phi(\mathbf{p})[k]} e^{-j\Delta\phi_{RX}} \tag{14}$$

where $\Delta\phi(\mathbf{p})[k]$ is the time series of PDOAs in (8), if $\mathbf{p} = \mathbf{p}_{TX}$. For DG/DPD, equation (14) is evaluated at every candidate emitter position. Since this evaluation over large chunks of the grid is computationally intensive, it can be accelerated by exploiting the massive parallel computing power of GPUs. To perform QDG, the computation in (14) is approximated by cross-correlating the pre-processed STFTs of $y_0[k]$ and $y_1[k]$:

$$\bar{S}(\mathbf{p}) = \sum_{m=0}^{M-1} \sum_{n \epsilon \bar{\mathbf{N}}_m} Y_0\big[m, n(\mathbf{p})[m]\big] Y_1^*\big[m, n(\mathbf{p})[m]\big] e^{-j\Delta\phi(\mathbf{p})[m]} e^{-j\Delta\phi_{RX}} \tag{15}$$

where $n(\mathbf{p})[m] = \lfloor(\Delta f_{LO} + f_d(\mathbf{p})[mL] + 0.5\, f_s)/\Delta f\rceil$ are the frequency bin indices of the FOAs in (5) with $f_d(\mathbf{p})[m] = f_{d,0}(\mathbf{p})[m] \approx f_{d,1}(\mathbf{p})[m]$, while $\Delta\phi(\mathbf{p})[m]$ is the time series of PDOAs as a function of the AOAs in (9), and $\bar{\mathbf{N}}_m \subseteq \{0, \dots, N-1\}$ denotes a subset of STFT frequency bins at the m-th time bin. As described in the next subsection, these subsets can retain meaningful bins, while discarding noisy ones.
The cross-correlation approximation (15) is evaluated over the maximum coherent integration time T underlying the STFT in (10). This coherent interval is upper bounded by the STFT time-frequency quantization error, which dominates over physical effects and the RF front end. On the contrary, common-mode errors of the orbit and LO stability are cancelled out between the two antennas. To increase the SNR beyond this bound, the cross-correlation shall be accumulated over many snapshots of I/Q samples, which span disjoint time intervals. These snapshots may come from repeated short-time acquisitions and/or multiple revisits of the receiver. The accumulation is done through a semi-coherent combination, which sums the magnitudes of the cross-correlations in different coherent intervals, where the FOAs and PDOAs parameterized over the same position. Assuming that the noise power is constant over the total semi-coherent integration time, the resultant sum is scaled to SNR according to:

$$\mathrm{SNR}(\mathbf{p}) = \frac{\sum_{i=0}^{N_{NC}-1}\left(|\bar{S}_i(\mathbf{p})|^2\, g(\mathbf{p})^2/g_{max}^2 - \sum_m^{M_i-1}\sum_{n\in\bar{\mathbf{N}}_m}\gamma_m(\mathbf{p})[m]\right)}{\sqrt{\sum_{i=0}^{N_{NC}-1}\left(\sum_m^{M_i-1}\sum_{n\in\bar{\mathbf{N}}_m}\gamma_m(\mathbf{p})[n]\right)^2}} \tag{16}$$

$$\gamma_m(\mathbf{p})[n] = 4\hat{\sigma}_n^4 E_w^2\left(\frac{\left|Y_0[m, n(\mathbf{p})[m]]\right|^2 g(\mathbf{p})}{2\hat{\sigma}_n^2 E_w\, g_{max}} + \frac{\left|Y_1[m, n(\mathbf{p})[m]]\right|^2 g(\mathbf{p})}{2\hat{\sigma}_n^2 E_w\, g_{max}} - 1\right) \tag{17}$$

where $g(\mathbf{p})$ denotes unitless coefficients that compensate for any spatial differences in the receiver visibility over the grid, by normalizing with respect to $g_{max} = \max_{\mathbf{p}} g(\mathbf{p})$. For example, unless the grid covers only the footprint of the overlapping main lobes of the antennas, where the gains have roughly a constant value, some positions may fall onto the nulls or the secondary lobes with consequent loss of SNR. Therefore, these coefficients compensate for angular differences by normalizing the patterns of the antennas across the positions.
The SNR defined in (16) is deflection metric that measures how many standard deviations the cross-correlation magnitude rises above noise when significant signal energy is present. It is useful for geolocating signal sources in the position domain, while it is not a physical metric usable for link budgets or comparable to thermal noise power.
It is worth noting that the presence of the phase offset $\Delta\phi_{RX}$ added to (8) does not reduce the SNR because it is constant.

In theory, any direct emitter geolocation in a single-receiver scenario relies on accurately knowing the waveform of the signal received, which is assumed in (Pojani et al., 2026; Tegedor et al., 2026; Ellis et al., 2020; Amar et al., 2008). This limitation prevents non-cooperative and unknown emitters from being geolocated when only one receiver is in visibility. The approximation introduced by QDG in (15) matches the history of Doppler shifts inside the FOAs at the receiver and it carries out carrier-phase interferometry with the PDOAs between the two antennas. Because the receiver is one, theoretically, this technique works effectively under the implicit assumption that the signal is a CW at the known transmit frequency $f_{TX}$. This would limit the generic signal model (2) to the special case with $s(t) = 1$. Extending the effectiveness of using a receiver with two antennas and QDG to generic narrowband waveforms is useful to tackle CWs at unknown carriers as well as swept-frequency modulations, such as periodic chirps, which might be partially unknown. For this purpose, the solution is expanding the search space by integrating over a sweep of transmit frequencies, which ideally sample the bandwidth $B_{TX}$. Given a set of trial frequencies **F**, we may expand (2) into:

$$\bar{S}(\mathbf{p}) = \sum_{f_{TX}\in\mathbf{F}} \bar{S}(\mathbf{p}, f_{TX}) \tag{18}$$

### 4.5 Single-Antenna Correlation Approximation

The special case of QDG with one individual antenna leads to the reformulate (15) as the correlation with a CW at position p, where the Doppler shifts are matched to the FOAs coming from arbitrary position vector **p**:

$$\bar{S}(\mathbf{p}) = \sum_{m=0}^{M-1}\sum_{n\in\bar{\mathbf{N}}_m} Y[m, n(\mathbf{p})[m]]\, e^{-j2\pi\tilde{\psi}(\mathbf{p})[m]} \tag{19}$$

where $\tilde{\psi}(\mathbf{p})[m]$ is the integral in time of the residual fractional FOA, i.e., $\varepsilon_f(\mathbf{p})$, after the STFT frequency bins compensates for the integer FOAs indexed $n(\mathbf{p})[m]$ over the time bins:

$$\widetilde{\psi}(\mathbf{p})[m] = \int_0^T \varepsilon_f(\mathbf{p}, t')dt', \qquad \varepsilon_f(\mathbf{p})[m] = \Delta f_{LO} + f_d(\mathbf{p})[mL] - n(\mathbf{p})[m]\Delta f - 0.5\, f_s \tag{20}$$

As opposed to the multi-antenna cross-correlation in (15), the maximum coherent integration time in (19) is further limited by the orbit errors and LO instability, which can exceed the quantization error of the STFT. To the extent possible, the SNR can be increased with semi-coherent combination by summing up the magnitudes of the correlations in different coherent intervals, which are short-time snapshots of I/Q samples. The resultant SNR expressed as a deflection metric is:

$$\mathrm{SNR}(\mathbf{p}) = \frac{\sum_{i=0}^{N_{NC}-1}\left(|\bar{S}_i(\mathbf{p})|^2\, g(\mathbf{p})/g_{max} - 2\hat{\sigma}_n^2 E_w \sum_m^{M_i-1} |\bar{\mathbf{N}}_m|\right)}{2\hat{\sigma}_n^2 E_w \sqrt{\sum_{i=0}^{N_{NC}-1}\left(\sum_m^{M_i-1}|\bar{\mathbf{N}}_m|\right)^2}} \tag{21}$$

**4.6 Computational Efficiency**

The use of the STFTs as look-up tables lowers the complexity to evaluate the cross-correlation in (15) compared to (14), because it eliminates the need to perform fractional delay interpolation and Doppler shift mixing onto many samples. This speeds up the grid search to the extent that chunks of positions can be processed sequentially through the cores of a central processing unit (CPU) instead of necessarily parallelizing the computation through a GPU.
In big O notation, the number of operations to pre-process the I/Q samples from a generic number A of antenna elements scales as $O(AMN \log N)$. Given a grid of size G, the number of operations to process the STFTs in the computation of the approximated cross-correlation (15) or correlation (19) scales as $O(AGM)$. Therefore, the complexity of QDG is $O(AMN \log N + AGM)$, whereas the complexity of the DG/DPD is $O(AGKQ)$, where Q is the cost of interpolating the expected fractional delays, Doppler frequencies and/or phase shifts over the I/Q samples. Therefore, the efficiency gain of QDG over DG/DPD becomes significant for large grids when $N\log N/G + 1 \ll LQ$ with $K \cong (M-1)L + N \sim ML$. This relation can be simplified to $LQ \gg 1$ with $G \gg 1$. In the following, $A = 2$, $L = 0.5\ N$, and N is set by the necessary time-frequency resolution of the STFT.

**4.7 Data Compression**

The STFT pre-processes the I/Q samples by projecting them into a time-frequency domain where the most important signal information is concentrated into a few bins. If the RF front ends properly amplify and are not saturated, it is likely that the majority of bins contain mostly noise. Therefore, even though (11) shows that the amount of data grows after pre-processing, this transform can be coupled with efficient algorithms of lossy data compression, which compress the STFT window by window, adapting to the received signals. One simple reference algorithm is de-noising: selectively filtering out the STFT bins with energy at or below the noise floor. This selection scales the subset $\bar{N}_m$ of unfiltered frequency bins over time to capture only the received signals with significant energy. The noise floor is identified with a threshold set according to (13) with an arbitrary false-alarm rate $P_{FA}$, which is the parameter that tunes the compression factor. A by-product of the compression is that the computation efficiency of the cross-correlation in (15) and correlation in (19) is increased, because the underlying coherent integration spans only the few time bins containing signal energy. Discarding noisy time-frequency bins translates into saving computations.
Compression in the time-frequency domain can be coupled with other classical methods of data compression of I/Q samples, such as reducing the bit depth though re-quantization (Tegedor et al., 2026) or downsampling after digital downconversion.

**4.7 Precision Bounds**

This sub-section outlines the general approach, steps, and error models to derive the Cramer-Rao lower bound (CRLB) of the QDG estimator. A comprehensive mathematical derivation is omitted here for brevity.
The CRLB quantifies the minimum achievable covariance of any locally unbiased estimator of the position of a signal-specific RF emitter. In this work, it is evaluated at each candidate position in the grid search. From the CRLB matrix, we calculate geolocation precision metrics in a local East-North-Up (ENU) reference frame, including horizontal and vertical errors and horizontal error-ellipse parameters. Under a local Gaussian approximation, percentile-based quantities such as the circular error probable (CEP) and the semi-major axis (SMA) of the error ellipse can be obtained from the horizontal covariance.
The CRLB matrix is the inverse of the Fisher information matrix (FIM), after accounting for nuisance parameters and for any prior or constraints. This matrix shall account for the models of meaningful information and error sources: the receiver state uncertainty, the measurement precision, the SNR, the geometry, and any prior or constraint on the altitude of the unknown position of the signal source.
The uncertainty on the receiver state is treated as an array of common nuisance parameters that are shared by all STFT time bins. These parameters are the receiver errors modeled as zero-mean random Gaussian perturbations. Their standard deviations characterize the position vector $\delta\mathbf{p}_{RX}$, the velocity vector $\delta\mathbf{v}_{RX}$, and the clock bias $\delta t_{RX}$. Errors on the receiver clock drift due to the LO instability are absorbed in the FOA measurement error model. As explained in (Ho et al., 2007), the prior receiver state covariance enters the CRLB through an augmented FIM as a nuisance block of the matrix. This prior is then removed with the Schur complement, yielding an effective FIM for the emitter position vector $\mathbf{p}_{\mathrm{TX}}$. The gradients that map the receiver state uncertainty into FOAs and PDOAs perturbations are computed with finite differences based on the orbit estimated on board.

The time series of FOAs at the STFT time bins matches the history of Doppler shifts of the signal received from the source position. To account for the contribution of FOAs to the FIM, the approach is to assume that STFT frequency bins behave like noisy FOA estimates with zero-mean Gaussian errors. With the support of experimental evidence shown in the error analysis of next section, these variables are set to have identical and SNR-independent standard deviation $\delta f$. The FIM contribution then accumulates the outer products of the spatial gradients of the FOA measurements in (5) with respect to $\mathbf{p}_{\mathrm{TX}}$ weighted by the inverse of the FOA error variance $\delta f^2$. This approach is a simplification, since the QDG does not necessarily estimate FOA explicitly. It actually performs a grid search where the FOA information is encoded directly in the amplitudes of the STFT bins indexed by the FOA time series. This simplified measurement model is most reasonable near good-SNR grid positions, which correspond to STFT bins that contain signal energy above the noise floor.

The time series of PDOAs at the STFT time bins explain the inter-antenna phase history of the signal received from the source position. To account for the contribution of PDOAs to the FIM, the approach is more complicated than for FOAs because it models the complex Gaussian noise affecting the phase of the inner product of the two STFTs. The FIM contribution then accumulates the outer products of the spatial gradients of the PDOA measurements in (8) with respect to $\mathbf{p}_{\mathrm{TX}}$ weighted by SNR of the STFT bins in (12). Moreover, with semi-coherent combination, in general, coherency can be assumed only inside each coherent interval but not across multiple intervals. This is modeled by introducing an unknown nuisance phase for each coherent interval and removing it through the Schur complement.

Although modeled separately, FOAs and PDOAs are statistically correlated as they are extracted from the same STFTs. With receiver uncertainty, the correct procedure is to build one augmented FIM containing both FOA and PDOA contributions to remove the receiver state prior only once. The result is joint effective FIM. It is worth noting that the FOA-only effective FIM can approximate well the joint effective FIM, when the PDOA contribution is weak or unreliable.

The unconstrained FIM quantifies the information of the three-dimensional emitter position vector $\mathbf{p}_{\mathrm{TX}}$. However, to constrain the search space of QDG to a two-dimensional grid of positions, the static signal source is assumed to lie at or near the Earth's surface. This is compatible with the ground search for terrestrial emitters carried out in this work. It means constraining the altitude component of $\mathbf{p}_{\mathrm{TX}}$ to the height of a global digital elevation model (GDEM) above the Earth ellipsoid, within the common datum used for instantiating the grid. Therefore, the altitude component of $\mathbf{p}_{\mathrm{TX}}$ is either treated as a prior or as a known height. In the former case, the FIM is soft-constrained with a height prior, which is modeled as a zero-mean Gaussian random variable with standard deviation $\delta h$. This inverse of this height variance $\delta h^2$ is added to the up-up element of the FIM with respect to the local ENU frame or, alternatively, it scales the local ellipsoidal unit vector added to the FIM in the Earth-centered Earth-fixed (ECEF) reference frame. The latter case represents a hard constraint, which is less conservative than the soft constraint because it simply removes the vertical degree of freedom from the FIM. In fact, any error in the GDEM height translates into a bias in the hard-constrained QDG estimator and an optimistic CRLB, which is assessed in (Ho et al., 1997).

## 5. JAMMER CANCELLATION AND ISOLATION

This section describes two methods to select and separate the individual signals that are received above the noise floor. These methods cancel or isolate the contribution of every signal to the position-domain SNR that is computed through QDG. They are used to recursively cancel one by one the sources that are visible and geolocated, in order of decreasing signal strength. This process can reveal the faint reception of low-SNR sources. The first emitters to be geolocated and cancelled are the strong and/or close ones, which are found at the positions of the primary peaks of SNR. As these are cancelled, secondary peaks of SNR might stand out and lead to the geolocation of other emitters, which are weak and/or distant and which would be otherwise overpowered by the first signal sources.

### 5.1 Beamforming and Null Steering

Consider the ideal case of a stationary RF emitter that is placed exactly at the position of the cross-correlation magnitude in (15) or the SNR in (16). By extracting the bins of the two STFTs that correspond to the predicted FOAs and rotating one STFT by the negative of the predictable PDOAs, these two STFTs are added coherently and constructively. This operation is equivalent to beamforming with a two-element antenna array: the gain of the array is steered towards the emitter position. The opposite operation of steering a null of the pattern in the same emitter direction is performed by flipping the sign of the second STFT. In reality, however, the FOAs and PDOAs are not perfectly parametrized for the emitter-receiver geometry; as a minimum, if the antenna array is not calibrated, the PDOAs are off by the constant phase offset $\Delta\phi_{\mathrm{RX}}$, which incorporates all phase biases. Nonetheless, this unknown phase offset can be estimated from the complex inner product of the bins, which inside the two STFTs correspond to the maximum cross-correlation magnitude. This estimation works by assuming that this maximum coincides with a signal source and is performed by averaging the phase over the times bins. This average can be weighed by the SNR in (12) of the STFTs at the two antennas as:

$$\Delta\widehat{\phi}_{RX} = -\angle\left(\sum_{m=0}^{M-1}\sum_{n\in\bar{\mathbf{N}}_m}\overline{SNR}[m,n]\frac{Y_0[m,n(\mathbf{p}_{max})[m]]Y_1^*[m,n(\mathbf{p}_{max})[m]]e^{-j\Delta\phi(\mathbf{p}_{max})[m]}e^{-j\Delta\phi_{RX}}}{|Y_0[m,n(\mathbf{p}_{max})[m]]Y_1^*[m,n(\mathbf{p}_{max})[m]]|}\right) \approx \Delta\phi_{RX} \quad (22)$$

where $\mathbf{p}_{max} = \mathrm{argmax}_{\mathbf{p}}\,\bar{S}(\mathbf{p})$ and, for example, $\overline{SNR}[m,n] = 0.5\,(SNR_0[m,n] + SNR_1[m,n])$. This processing can be summarized as calibrating the antenna array "on the fly" with respect to a signal coming from the source geolocated at $\mathbf{p}_{max}$. Once the array is calibrated, beams and the nulls of the antenna patterns can be steered in the coherent sum of the STFTs. To cancel the energy received from any signal source at the position vector **p**, we may define this coherent sum as:

$$\Sigma(\mathbf{p})[m,n] = 0.5\left(Y_0[m,n] - \zeta[m,n]\,Y_1[m,n]e^{j\Delta\phi(\mathbf{p})[m]+j\Delta\widehat{\phi}_{RX}}\right) \quad (23)$$

where the coefficient $\zeta[m,n] = |Y_0[m,n]|/|Y_1[m,n]|$ normalizes the amplitudes of the STFT bins. After cancellation, the sum of STFTs in (10) can be re-used further for QDG to compute the single-antenna SNR in (19), which will be ideally free from contributions of the cancelled signal source.
As a method for signal cancellation, null steering is inherently narrowband and sensitive to errors in the phase compensation and in the amplitude normalization across the antennas. The effectiveness degrades significantly with errors related to the receiver PVT and STFT quantization. Moreover, it is challenged when every antenna element has an independent RF chain, with separate low-noise amplifier (LNA) and automatic gain control (AGC). Finally, the number of nulls is upper bounded to the number of antenna elements minus one, which, with two elements, allows for the cancellation of only one signal.

### 5.2 Time-Frequency Filtering

A simpler and more flexible method for signal isolation/cancellation than beamforming or null steering is filtering in the time-frequency domain. This consists in multiplying the STFTs with a binary mask that selects the bins to isolate or discards the bins to cancel, before the filtered STFTs are used to compute the approximated correlation or cross-correlation in the position domain. For example, the following STFT is filtered to isolate the received signal with bandwidth $B_{TX}$ of a source through the element-wise product:

$$\widetilde{Y}[m,n] = Y[m,n] \odot H(\mathbf{p})[m,n], \qquad H(\mathbf{p})[m,n] = \begin{cases} 1 & \text{if } |n - n(\mathbf{p})[m]| \le 0.5\,B_{TX} \\ 0 & \text{if } |n - n(\mathbf{p})[m]| > 0.5\,B_{TX} \end{cases} \quad (24)$$

where $H(\mathbf{p})$ is the binary mask that sets the emitter position to **p**. This method does not require multiple antennas and array calibration, because it is based on the energy of the signals and not their relative phase. Moreover, it is not limited in the number of signals that can be cancelled or isolated.
By inverting the filtered STFT, it is possible to recover the I/Q samples that contain the signal isolated or that are free from the signals cancelled. Time-frequency filtering and inversion find applications in the mitigation of interference entering GNSS receivers and in the two-step geolocation of multiple jammers with overlapping waveforms in (Absdoush et al., 2017; Pojani et al., 2018).

## 6. EXPERIMENTAL RESULTS

This section presents the data and the results of an experiment of GNSS RFI geolocation from LEO, which was carried out with OPS-SAT PRETTY in occasion of Jammertest 2025. The coordination of the satellite operations for the acquisition and downlink of the datasets collected and the ground operations for the transmission of GNSS RFI were made possible by the collaboration of the Joint Research Centre (JRC) of the European Commission, ESA, the Technical University (TU) of Graz, Testnor, and the Norwegian Defense Research Establishment (FFI). The team at JRC later devised QDG and post-processed the data collected through this technique to produce the results, which are shown below.

### 6.1 Experiment Setting

The work presented draws from the dataset collected on September 12th during a dedicated test session made in Bleik, near Andøya, Norway, which occurred a few days before the official start of Jammertest 2025. This collection was part of a campaign that made other collections in the days of the event from September 15th to 19th.

On this day, the transmissions of GNSS RFI from the test jammer were planned to match four passes of OPS-SAT PRETTY, which were predicted to have good visibility of the transmission site. Likewise, the satellite was scheduled to acquire data during the same transmission windows. Details about the OPS-SAT PRETTY satellite and the collection strategy may be found in (Tegedor et al., 2026; Zeif et al., 2020). The test was conducted with the so-called "Porcus Major" high-power jammer, which is developed by the FFI. This test jammer was placed statically on ground around 69.2852899N 16.0013001E, from where it was emitting 50 W of effective isotropic radiated power (EIRP) in the form of a CW, simultaneously on the L1 and L5 carriers.
Out of the four passes, the results presented focus on the most favorable one, which lasted from 11:31:52 UTC to 11:44:52 on September 12th, with a maximum elevation of 50° above Bleik. In this 12-min window, the satellite made 27 acquisitions of 1 s each, every 9 s, storing about 1 GB of I/Q samples with both antennas in L5 alone. The samples were accompanied with logs of the on-board GNSS receiver and the attitude determination and control system (ADCS). The downlink of the dataset for the whole campaign, i.e., ~8 GB, occupied the ground station at TU Graz for the weeks following the experiment.
Fig. 1 and Fig. 2 plot the estimated track of OPS-SAT PRETTY and the estimated position precision, respectively, as reported by the on-board GNSS receiver during the most favorable pass. Fig. 3 shows the NIC registered by the local air traffic on the same day through the JRC's internal tool ADSBMon, which aggregates and visualizes a repository of open-source and commercial ADS-B data over Europe. Both the GMSS receiver log and the NIC exhibit evidence of RFI for assets flying above Andøya.

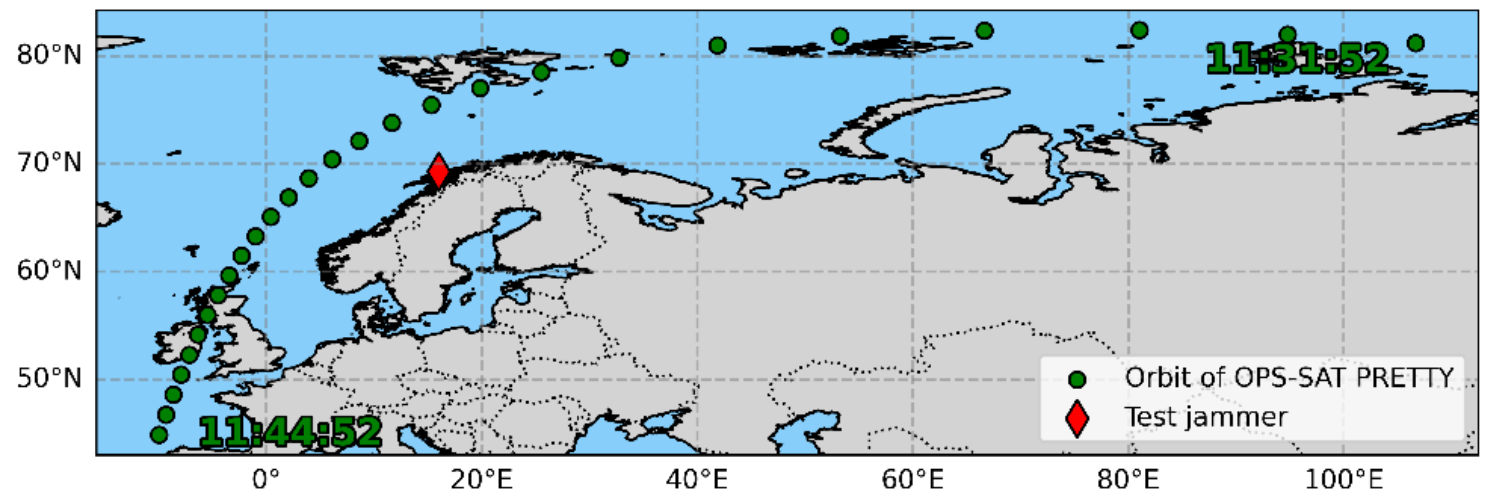

**FIGURE 1** Position of OPS-SAT PRETTY every 30 s during a pass of Sept. 12th based on the on-board GNSS receiver.

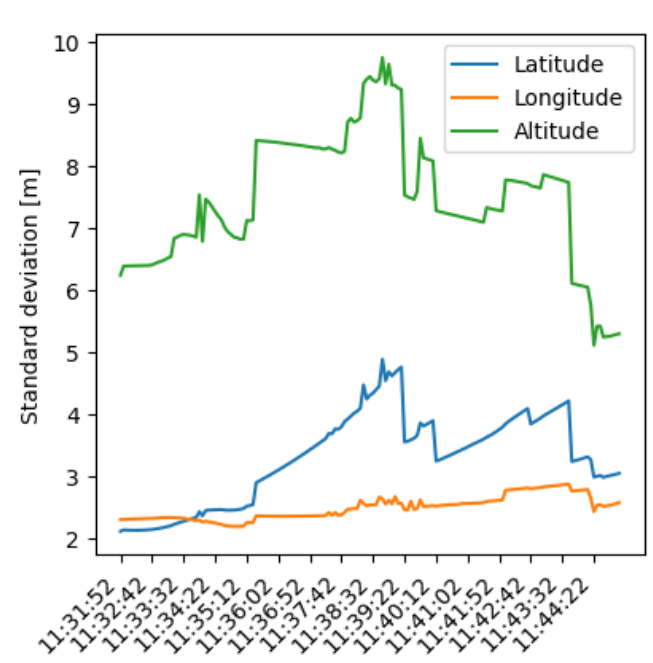

**FIGURE 2** Reported precision of the on-board GNSS receiver during the pass.

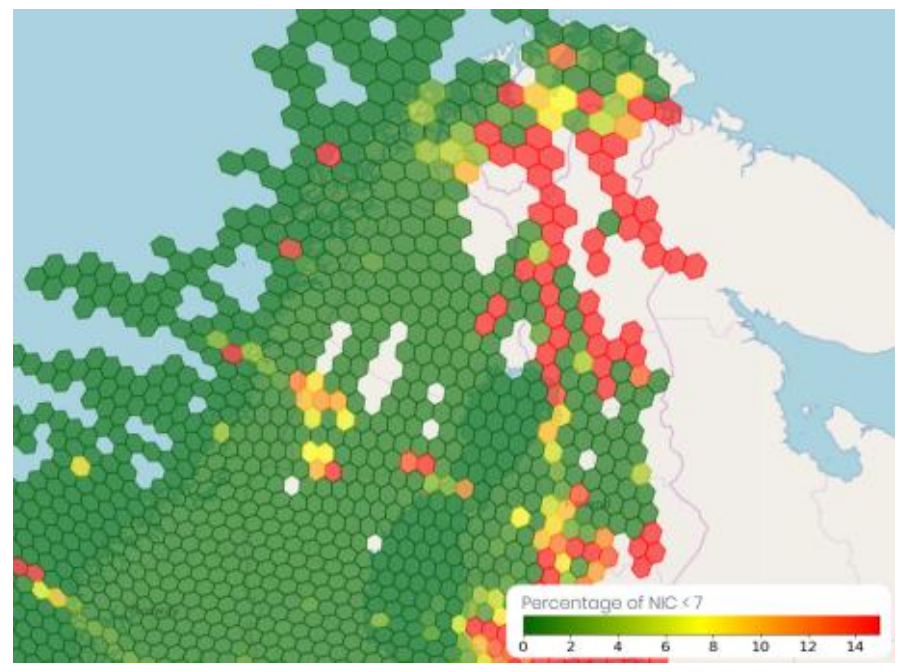

**FIGURE 3** Anomalies in NIC registered over Scandinavia on the same day.

### 6.2 Error Sources

OPS-SAT PRETTY is designed for GNSS-R and repurposed for RFI monitoring in this experiment. The LO is not disciplined by the GNSS receiver. A constant frequency offset $\Delta f_{LO} \approx 8.11$ kHz is estimated at L5 and compensated. This estimate is obtained by comparing the exact FOAs expected from the true position of the jammer with the nearby FOAs, which lie at the bins of the STFT where the SNR is maximum. After compensation, the residual error in frequency due to STFT quantization, LO, orbit, etc. is estimated to be $\delta f \approx 8$ Hz rms (see Fig. 4) with a rate of ~3 Hz/s rms.
The antenna array baseline is supposed to be kept in line with the along-track direction. On another pass of the same day, the comparison between the velocity estimated by the GNSS receiver and the quaternions from the ADCS show an attitude error below 1° rms.
Since the antenna array is not calibrated, an unknown phase offset exists between the two patch elements. After QDG, this offset is confirmed to be roughly constant at sufficient SNR, as shown in Fig. 5. It is estimated to be $\Delta\phi_{RX} \approx 148.1°$ by using (22) with the semi-coherent combination of the 27 acquisitions at the position of maximum cross-correlation magnitude or SNR, namely at the true jammer position. This estimate is used for null steering in (23).
The baseline spans $d = 105$ mm, which is about 0.41 $\lambda$ at L5. This means that the PDOAs do not wrap around the baseline but have a range cramped to ±148°. Therefore, the PDOA is more sensitive to errors for AOAs from the sides of the array and more precise for AOAs from boresight. Luckily, the expected AOAs of the test jammer range from 26° to 156°. The resultant PDOA error is estimated to be ~7° rms.

For simplicity, the receiver state covariance is assumed as diagonal and constant over time. By averaging the precision estimates plotted in Fig. 2 in the ENU reference frame, the positions standard deviations are calculated to be $\delta\mathbf{p}_{\mathrm{RX}} = [2.5, 3.4, 7.5]$ m. The other priors are set according to order-of-magnitude assumptions for single-frequency LEO receivers: the velocity deviations are $\delta\mathbf{v}_{\mathrm{RX}} = [5.0, 5.0, 5.0]$ cm/s and the clock bias deviation is $\delta t_{\mathrm{RX}} = 100$ ns.
Since the test jammer is placed on ground, the altitudes of the grid positions are set at heights that are interpolated from the global digital elevation model (GDEM) in (Danielson et al., 2010), which is regridded at 0.0625° to occupy just 26 MB. The coarse resolution of this model over the undulated terrain of Andøya causes a height uncertainty, which is evaluated to be $\delta h \approx 300$ m. Therefore, an equivalent prior is added to constrain the FIM of the QDG estimator to the GDEM.

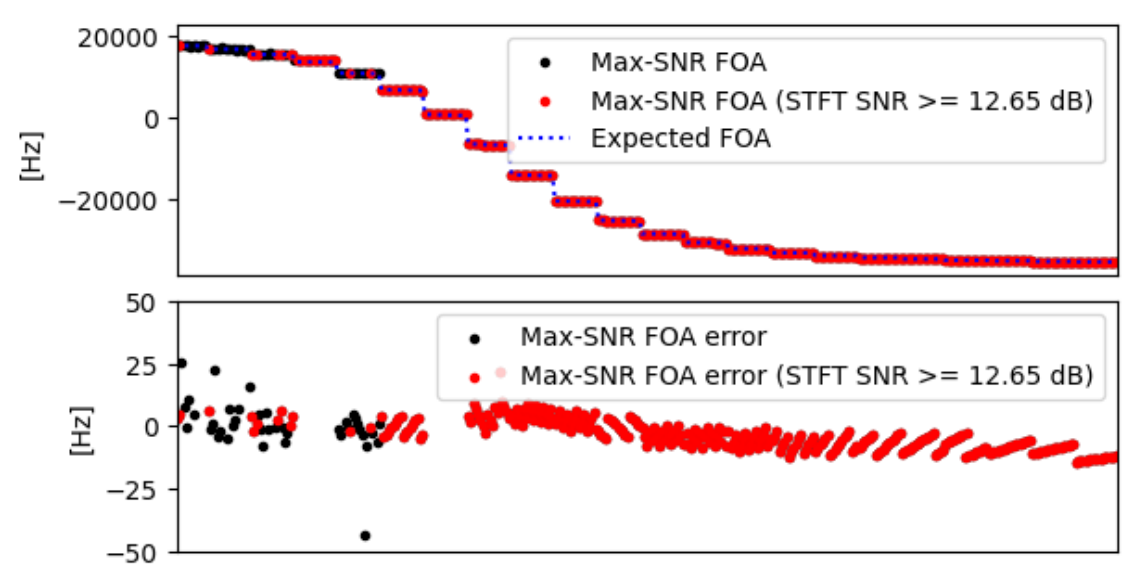

**FIGURE 4** Residual frequency errors after compensation over STFT time bins and acquisitions.

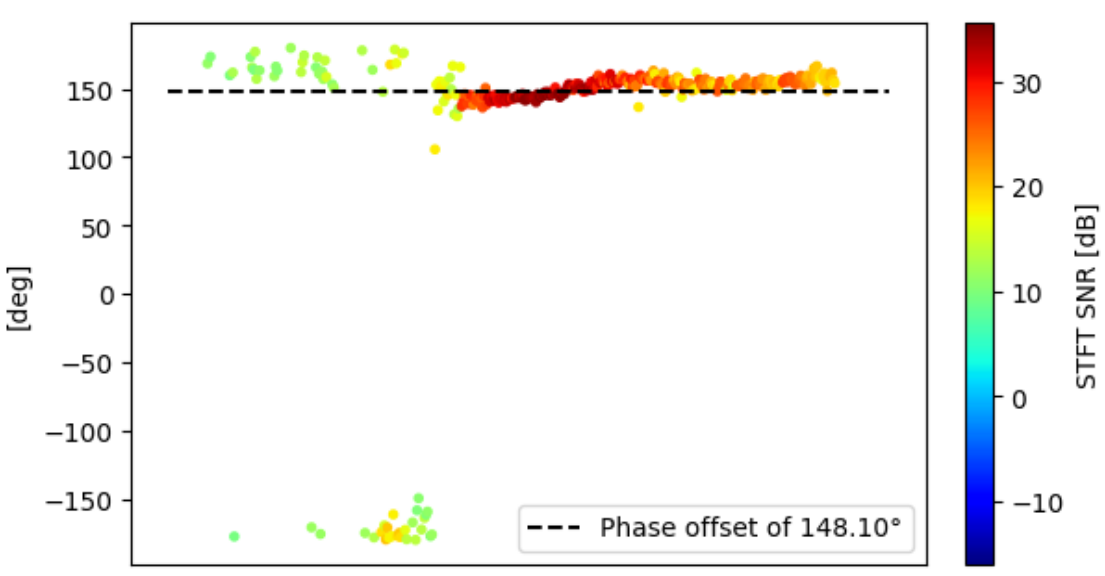

**FIGURE 5** Phase offset estimation over STFT time bins and acquisitions.

### 6.3 Control of Data Compression

The I and Q samples were acquired at 5 Msps with an effective depth of 12 bits and stored as 16 bits. At the pre-processing stage, the STFT of these samples are computed using Hann windows with 50% overlap and $N = 2^{19}$, which leads to resolutions of $\Delta f \approx 10$ Hz and $\Delta t \approx 52$ ms, and is optimized for narrowband waveforms as formalized in (Abdoush et al., 2019a). The memory sizes occupied by the inner products of two STFTs across the 27 acquisitions are indicated in Table 1. Each unfiltered value of this product is stored as a 64-bit complex floating-point and accompanied by a 32-bit unsigned integer for the index corresponding to the relevant time and frequency bin. Hypothetically, in this experiment, these memory sizes provide an indication of the compressed data that can be downlinked to perform QDG in post-processing or processed in orbit with an OBC. Two compression methods are combined. The first is the classical method of reducing the bit depth of the I/Q samples, while minimizing the quantization losses. The second method is filtering out the time-frequency bins with energy at or below the noise threshold in (13), which can be tuned with the false-alarm rate $P_{\mathrm{FA}}$. The table also reports the noise estimates across the two antennas and the sensitivities of the compression relative to the ±3 dB variation in the noise estimates.
It is noted that the extremely high compression factors (e.g., 99.9%) are achievable with the dataset under analysis, because the test jammer and any other emitter in sight during the satellite pass were transmitting CWs. Consequently, the STFT bins with energy above noise are only ~0.08% of the total number of bins. High compression ratios are still likely with frequency-swept or frequency-hopped modulations, provided that the analysis window chosen to set the STFT time-frequency resolution is suitable to capture the instantaneous frequency of the waveform.

**TABLE 1** Memory size of the dataset needed to finalize QDG after pre-processing and compression.

| * No compression | | | | False-alarm rate | | | | |
|---|---|---|---|---|---|---|---|---|
| ($\hat{\sigma}_n^2 \mp 3$ dB) | | **I/Q** | $\mathbf{2\hat{\sigma}_n^2 E_w}$ | **1*** | **0.1** | **0.01** | **0.001** | **0.0001** |
| **Bit depth** | **2 bit** | 135 MB | -10.05 dBFS | 4076.9 MB | 19.10 MB (161.6, 1.01) | 2.02 MB (34.87, 0.17) | 0.52 MB (9.25, 0.08) | 0.24 MB (3.04, 0.05) |
| | **4 bit** | 270 MB | -10.21 dBFS | 4076.9 MB | 23.56 MB (181.3, 1.38) | 2.72 MB (41.94, 0.24) | 0.70 MB (11.75, 0.10) | 0.31 MB (4.04, 0.06) |
| | **8 bit** | 540 MB | -10.43 dBFS | 4076.9 MB | 25.64 MB (191.4, 1.51) | 2.98 MB (45.27, 0.26) | 0.76 MB (12.86, 0.11) | 0.33 MB (4.44, 0.07) |
| | **16 bit* (12 bit)** | 1080 MB | -10.43 dBFS | 4076.9 MB | 25.48 MB (190.7, 1.50) | 2.96 MB (45.03, 0.25) | 0.76 MB (12.78, 0.11) | 0.33 MB (4.41, 0.07) |

### 6.4 Geolocation of the Test Jammer

The results presented in this sub-section are based on processing 0.76 MB of data that store the inner products of the two STFTs, according to the configuration underlined in Table 1. This configuration compresses the I/Q samples by re-quantizing to 8 bits and by de-noising with $P_{\mathrm{FA}} = 0.001$. By integrating these compressed data, the SNR (16) is computed over grids of WGS84 positions by processing the FOA/PDOA-based cross-correlation with the two antennas. This SNR semi-coherently combines all the 27 acquisitions, each of which is coherently integrated for the full 1-s time in (15). The two grids in Fig. 6 and in Fig. 7 and 8 are centered on the true position of the test jammer in Bleik. Both count $G \approx 125000$ positions, although one spans a wide region with a coarse resolution, while the other one focuses on the local area of Andøya with a fine resolution. Fig. 9 plots the precision upper bound in the form of the isotropic MSA at 95$^{th}$ percentile (MSA95) to capture the error ellipsis variation in space depending mainly on geometry and SNR. This metric is calculated from the CRLB of the QDG estimator as described in Section 3. The figure shows that the estimation is geometrically very imprecise along the ground track of the satellite orbit, and it shows the beneficial impact of the STFT bins with high SNR. The maximum peak of the SNR in (15) is at the center in both the wide and the narrow grids, thus falling within the 0.5-km² area around the ground truth. The MSA95 at this maximum is ~900 m around the test jammer. At this position, the error ellipse is very elongated, since the 95$^{th}$-percentile semi-minor axis is ~50 m, and it is tilted 33° South to East. The altitudes of the candidate positions are set by interpolating the GDEM, which introduces evident errors in height due to the terrain undulation (see Fig. 8 vs. Fig. 10).

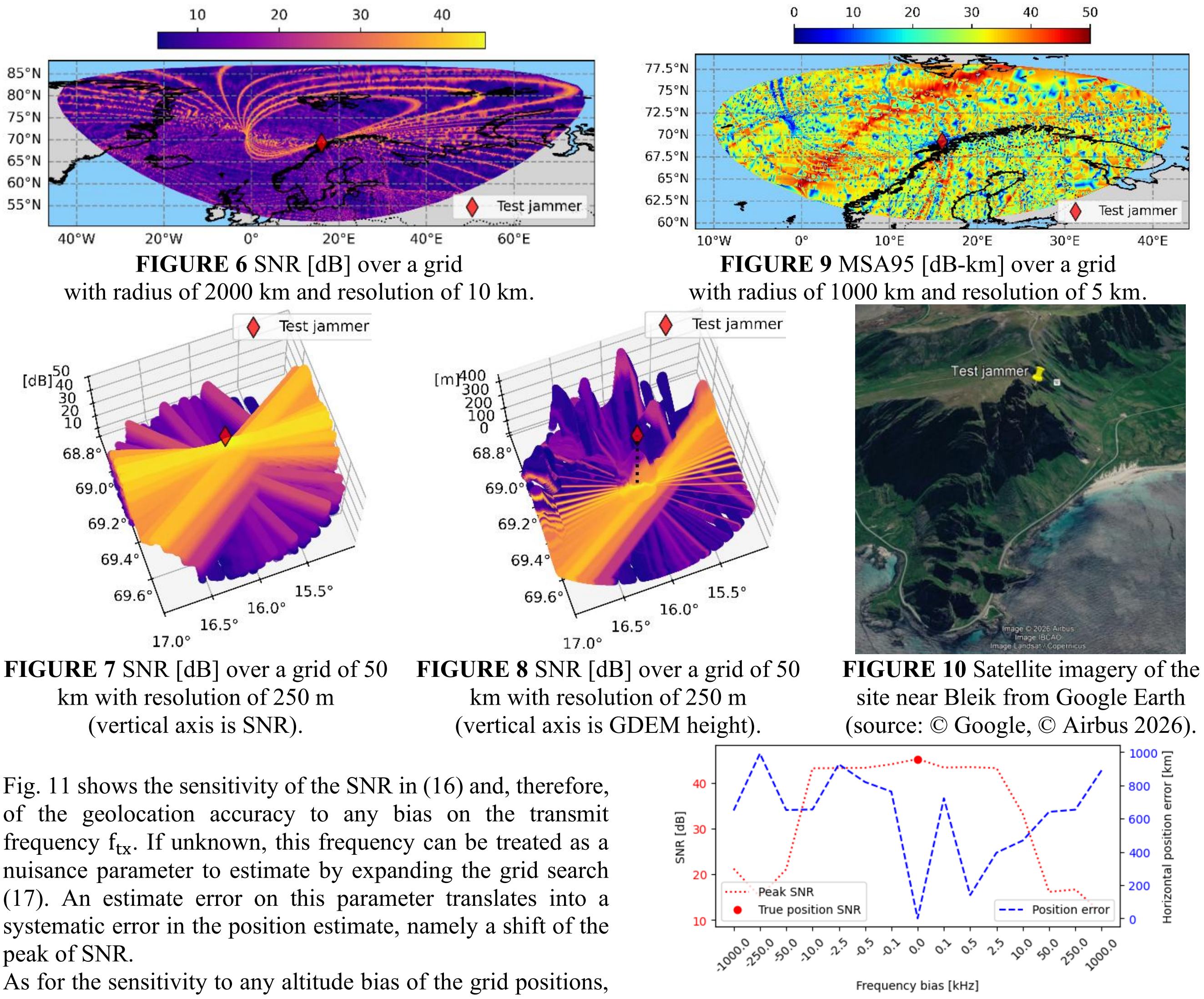

**FIGURE 6** SNR [dB] over a grid with radius of 2000 km and resolution of 10 km.

**FIGURE 9** MSA95 [dB-km] over a grid with radius of 1000 km and resolution of 5 km.

**FIGURE 7** SNR [dB] over a grid of 50 km with resolution of 250 m (vertical axis is SNR).

**FIGURE 8** SNR [dB] over a grid of 50 km with resolution of 250 m (vertical axis is GDEM height).

**FIGURE 10** Satellite imagery of the site near Bleik from Google Earth (source: © Google, © Airbus 2026).

Fig. 11 shows the sensitivity of the SNR in (16) and, therefore, of the geolocation accuracy to any bias on the transmit frequency $f_{tx}$. If unknown, this frequency can be treated as a nuisance parameter to estimate by expanding the grid search (17). An estimate error on this parameter translates into a systematic error in the position estimate, namely a shift of the peak of SNR.

As for the sensitivity to any altitude bias of the grid positions, this is accounted for in the prior $\delta h$ on the height error of the GDEM and, as such, is captured by an enlarged error ellipse.

**FIGURE 11** SNR [dB] vs. horizontal position error [m] over 5-km resolution grid for different biases in frequency.

### 6.5 Cancellation of the Test Jammer

The results in this sub-section present outcomes of the cancellation of the test jammer from the received signals. Fig. 12 shows the SNR in (21) computed after steering a null of the antenna array towards Bleik. This is achieved by processing the single-antenna and FOA-based correlation in (19) with the coherent sum of the STFTs after calibration, which injects in (15) the constant phase offset estimated with respect to the test jammer itself. Fig. 13 shows the SNR of the two-antenna and FOA/PDOA-based cross-correlation after filtering out the time-frequency bins of the STFTs that correspond to the position of the test jammer. This second method clearly delivers a more effective cancellation with less residual energy. The cancellation of the strong signal from Bleik reveals more energy that arrives with low SNR and that can potentially lead to other emitters. The areas in which these distant emitters seem to be located are traceable to regions known to be hotpots of RFI, which leak inside Europe affecting especially the Baltic countries. In fact, the extent of impact of GNSS RFI on these regions is plain to see in the NIC degradations registered on the same day and visualized through JRC's ADBMon tool in Fig. 14. By comparison, the energy seen by OPS-SAT PRETTY from 2000 km away and stored in less than 1 MB of data arguably conveys a more accurate insight into the origins of these RFI emissions (see Fig. 15) than the aggregation of GNSS performance proxy indicators from ADS-B data.

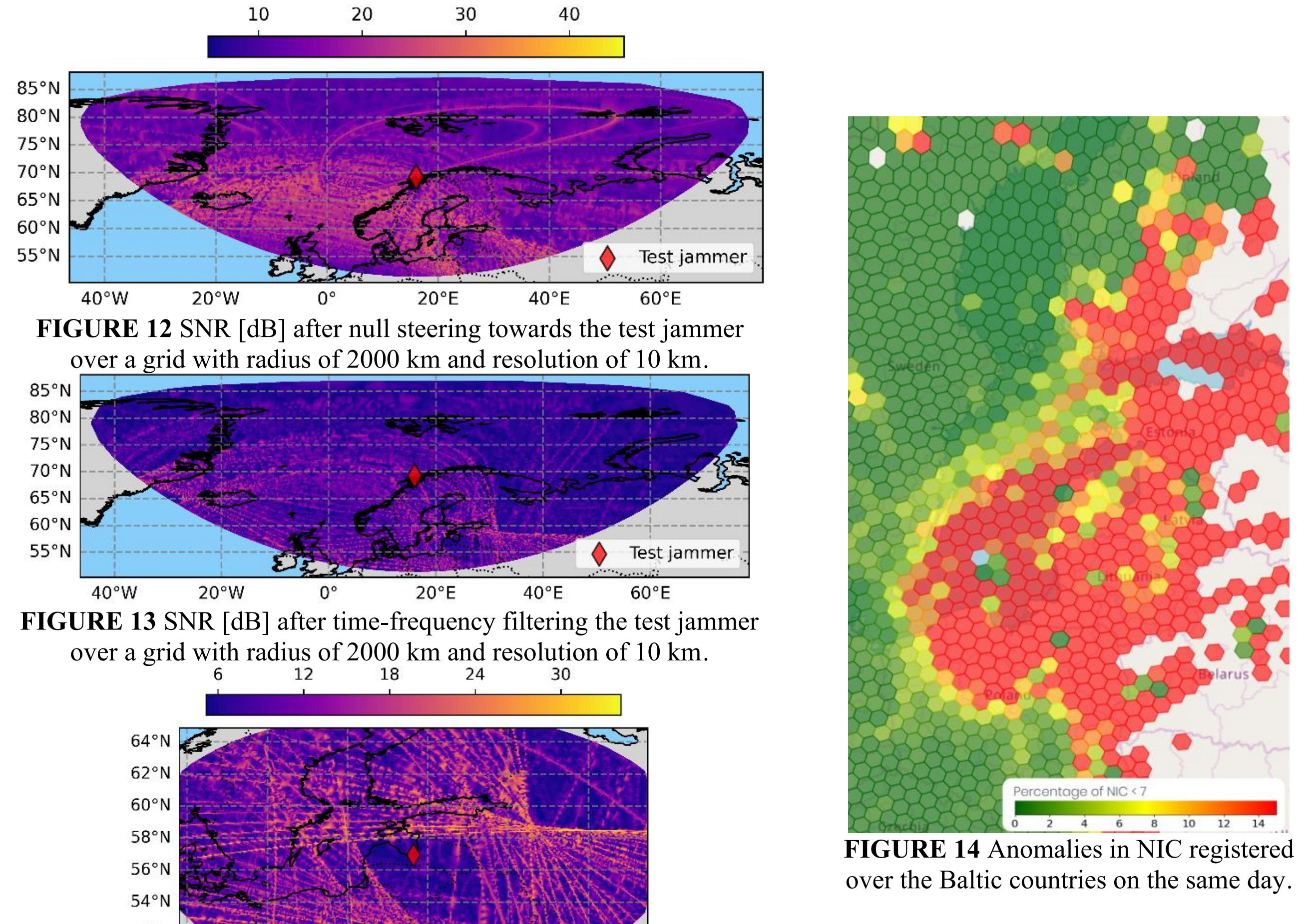

**FIGURE 12** SNR [dB] after null steering towards the test jammer over a grid with radius of 2000 km and resolution of 10 km.

**FIGURE 13** SNR [dB] after time-frequency filtering the test jammer over a grid with radius of 2000 km and resolution of 10 km.

**FIGURE 14** Anomalies in NIC registered over the Baltic countries on the same day.

**FIGURE 15** SNR [dB] after time-frequency filtering the test jammer over a grid with radius of 1000 km and resolution of 5 km, and centered in Riga (for reference).

## 7. CONCLUSION

This paper presents a comprehensive analysis and performance evaluation of QDG based on the dataset collected by OPS-SAT PRETTY in one pass over Jammertest 2025. The compression factors achievable with this technique are evaluated for various combinations of re-quantization and time-frequency de-noising. In the configuration used, 1 GB of I/Q samples across both patch antennas is compressed into less than 1 MB of data. This data is the inner products of the STFTs for the two antennas and 27 acquisitions of 1 s each. By integrating the bins with meaningful energy of this STFT product, QDG can pinpoint the test jammer in Bleik emitting 50 W of CW at L5 with accuracies within 0.5 km$^2$. This accuracy attains the theoretical precision upper bound, which is derived as part of the analysis. The same compressed data can be further processed to sequentially isolate or cancel signals through either beamforming and null steering or time-frequency filtering. The latter method is more flexible and does not require multiple antennas and antenna array calibration. When applied to the compressed data, this method is shown to enable the geolocation of low-SNR signals from sources that are weak or far from the satellite. In fact, after cancelling the CW of the test jammer, correlation peaks of other RFI sources, which would be otherwise hidden, rise above the noise. These emitters can be traced to well-known GNSS RFI hotspots, which are near the Baltic countries, a good 2000 km away from the satellite.

The controllable complexity and the flexibility of the ensemble of signal processing algorithms that compose QDG makes this technique usable on platforms with limited communication bandwidth and computational power. This is the case of LEO satellites constrained in SWaP, which usually have low-power OBCs and low-capacity downlinks. QDG has the potential to enable the use of these space assets to compute quick looks of the RF energy on the Earth's surface, which, in the GNSS bands, accurately geolocate RFI hotspots and sources. The processing behind QDG is feasible in near real time, either in orbit or, after downlinking small amounts of compressed data, on ground. The efficiency achieved with this technique relaxes the requirements on the platforms that can potentially deliver persistent and low-latency monitoring of GNSS RFI from LEO. This unlocks also the opportunistic exploitation of single and repurposed satellites, as demonstrated with OPS-SAT PRETTY.

## 8. FUTURE WORK

This work has established the mathematical framework and demonstrated the potential of QDG with real data from a satellite in LEO. The natural next steps are to validate the technique in a multi-satellite scenario and to implement the relevant signal processing algorithms for real-time compression and geolocation into a system that embeds a low-power CPU or GPU.